\documentclass[11pt]{article}

\usepackage[english]{babel}
\usepackage[latin1]{inputenc}
\usepackage[T1]{fontenc}

\usepackage{fullpage}
\usepackage{times}
\usepackage{amssymb,amsfonts,amsmath}
\usepackage{theorem}
\usepackage[sort]{cite}
\usepackage{paralist}

\newtheorem{theorem}{Theorem}
\newtheorem{proposition}[theorem]{Proposition}
\newtheorem{lemma}[theorem]{Lemma}

{\theorembodyfont{\rmfamily}
  \newtheorem{definition}[theorem]{Definition}}
{\theorembodyfont{\rmfamily}}
{\theorembodyfont{\rmfamily}\newtheorem{remark}[theorem]{Remark}}
\newenvironment{proof}%
   {\begin{trivlist}\item[]\emph{Proof:}}%
   {\end{trivlist}}
\newenvironment{proofof}[1]%
   {\begin{trivlist}\item[]\emph{Proof of #1:}}%
   {\end{trivlist}}

\newcommand\squareforqed{\hbox{$\Box$}}
\newcommand\qed{\ifmmode\squareforqed\else{\unskip\nobreak\hfil
   \penalty50\hskip1em\null\nobreak\hfil\squareforqed
   \parfillskip=0pt\finalhyphendemerits=0\endgraf}\fi}

\newcommand{\ep}{\textsc{ExPos}}

\newcommand{\np}{\textsc{NP}}
\newcommand{\logspace}{\textsc{LogSpace}}

\newcommand{\cproblem}[3]%
{\begin{trivlist}
  \item[]%
    \textbf{Problem:} \textsc{#1}\\
    \textit{Input:} #2\\
    \textit{Question:} #3
  \end{trivlist}%
}

\newcommand{\N}{\mathbb{N}}

\renewcommand{\phi}{\varphi}

\newcommand{\csp}{\ensuremath{\mathrm{CSP}}}

\newcommand{\card}[1]{\lvert#1\rvert}

\title{\bfseries Complexity of Existential Positive First-Order Logic}

\author{%
  Manuel Bodirsky,
  Miki Hermann\\
  LIX (UMR CNRS 7161), {\'E}cole Polytechnique, 91128 Palaiseau, France\\
  \{bodirsky, hermann\}@lix.polytechnique.fr
  \and
  Florian Richoux\thanks{This work was done
    during the PhD studies of the third author at {\'E}cole
    Polytechnique.}\\
  JFLI, CNRS - University of Tokyo, Japan\\
  richoux@jfli.itc.u-tokyo.ac.jp
}

\date{}

\begin{document}

\maketitle

\begin{abstract}
  Let~$\Gamma$ be a (not necessarily finite) structure with a finite
  relational signature.  We prove that deciding whether a given
  existential positive sentence holds in~$\Gamma$ is in $\logspace$ or
  complete for the class $\csp(\Gamma)_\np$ under deterministic
  polynomial-time many-one reductions. Here, $\csp(\Gamma)_\np$ is the
  class of problems that can be reduced to the \emph{constraint
    satisfaction problem} of~$\Gamma$ under \emph{non-deterministic}
  polynomial-time many-one reductions.

  \medskip

  \noindent%
  Key   words:    Computational   Complexity,   Existential   Positive
  First-Order Logic, Constraint Satisfaction Problems
\end{abstract}

\section{Introduction}

We study the computational complexity of the following class of
computational problems.  Let~$\Gamma$ be a structure with finite or
infinite domain and with a finite relational signature.  The
model-checking problem for existential positive first-order logic,
parametrized by~$\Gamma$, is the following problem.

\cproblem{$\ep(\Gamma)$}%
{An existential positive first-order sentence~$\Phi$.}%
{Does~$\Gamma$ satisfy~$\Phi$?}

\noindent%
\emph{Existential positive first-order formula} over~$\Gamma$ are
first-order formulas without universal quantifiers, equalities, and
negation symbols, and formally defined as follows:
\begin{compactitem}[-]
\item if $R$ is a relation symbol of a relation from $\Gamma$ with
  arity~$k$ and $x_1, \ldots, x_k$ are (not necessarily distinct)
  variables, then $R(x_1, \ldots, x_k)$ is an existential positive
  first-order formula (such formulas are called \emph{atomic});
\item if~$\varphi$ and~$\psi$ are existential positive first-order
  formulas, then $\varphi \land \psi$ and $\varphi \lor \psi$ are
  existential positive first-order formulas;
\item if~$\varphi$ is an existential positive first-order formula with
  a free variable~$x$ then $\exists x. \varphi$ is an existential
  positive first-order formula.
\end{compactitem}
An \emph{existential positive first-order sentence} is an existential
positive first-order formula without free variables.

Note that we do not allow the equality symbol in the existential
positive sentences; this only makes our results stronger, since one
might always add a relation symbol $=$ for the equality relation to
the signature of $\Gamma$ to obtain the result for the case where the
equality symbol is allowed.  Also note that adding a symbol for
equality to $\Gamma$ might change the complexity of $\ep(\Gamma)$.
Consider for example $\Gamma := (\N;\neq)$; here, $\ep(\Gamma)$ can be
reduced to the Boolean formula evaluation problem (which is known to
be in $\logspace$) as follows: atomic formulas in~$\Phi$ of the form
$x \neq y$ are replaced by \emph{true}, and atomic formulas of the
form $x \neq x$ are replaced by \emph{false}. The resulting Boolean
formula is equivalent to true if and only if~$\Phi$ is true
in~$\Gamma$.  However, the problem $\ep(\Gamma')$ for $\Gamma' :=
(\N;\neq,=)$ is NP-complete. Similar examples exist over finite
domains.

The \emph{constraint satisfaction problem} $\csp(\Gamma)$ for~$\Gamma$
is defined similarly, but its input consists of a \emph{primitive
  positive} sentence, that is, a existential positive sentence without
disjunctions.  Constraint satisfaction problems frequently appear in
many areas of computer science, and have attracted a lot of attention,
in particular in combinatorics, artificial intelligence, finite model
theory and universal algebra; we refer to the recent collection of
survey articles on this subject~\cite{Vollmer-08}.  The class of
constraint satisfaction problems for infinite structures $\Gamma$ is a
rich class of problems; it can be shown that for every computational
problem there exists a relational structure $\Gamma$ such that
$\csp(\Gamma)$ is equivalent to that problem under polynomial-time
Turing reductions~\cite{BodirskyG-08}.

In this paper, we show that the complexity classification for
existential positive first-order sentences over infinite structures
can be reduced to the complexity classification for constraint
satisfaction problems.  For finite structures~$\Gamma$, our result
implies that $\ep(\Gamma)$ is in $\logspace$ or $\np$-complete.  The
$\logspace$-solvable cases of $\ep(\Gamma)$ are in this case precisely
those relational structures~$\Gamma$ with an element~$a$ such that all
non-empty relations in~$\Gamma$ contain the tuple $(a,\ldots, a)$; in
this case, $\ep(\Gamma)$ is called \emph{$a$-valid}.  Interestingly,
this is no longer true for infinite structures~$\Gamma$.  To see this,
consider again the structure $\Gamma := (\N;\neq)$, which is clearly
not $a$-valid, but in $\logspace$ as we have noticed above.

A universal-algebraic study of the model-checking problem for finite
structures $\Gamma$ and various other syntactic restrictions of
first-order logic (for instance positive first-order logic) can be
found in~\cite{Martin-08}.

A preliminary version of this article appeared in~\cite{BHR-09}. The
present version differs in that the main proof has been simplified and
now also works without the relation symbol for equality; moreover,
Proposition~\ref{prop:loc-refut} and Section~\ref{sect:functions} have
been added.

\section{Main Result}

We write $L \leq_m L'$ if there exists a deterministic polynomial-time
many-one reduction from~$L$ to~$L'$.

\begin{definition}[from \cite{LadnerLS-75}]
  A problem~$A$ is \emph{non-deterministic polynomial-time many-one
    reducible} to a problem~$B$ ($A \leq_\np B$) if there is a
  nondeterministic polynomial-time Turing machine~$M$ such that $x \in
  A$ if and only if there exists a computation of $M$ that outputs $y$
  on input $x$, and $y \in B$.  We denote by $A_\np$ the smallest
  class that contains~$A$ and is downward closed under $\leq_\np$.
\end{definition}

Observe that $\leq_\np$ is transitive~\cite{LadnerLS-75}.  To state
the complexity classification for existential positive first-order
logic, we need the following concept.  The $\Gamma$-\emph{localizer}
$F(\psi)$ of a formula~$\psi$ is defined as follows:
\begin{compactitem}
\item $F(\exists x. \psi) = F(\psi)$
\item $F(\varphi \land \psi) = F(\varphi) \land F(\psi)$
\item $F(\varphi \lor \psi) = F(\varphi) \lor F(\psi)$
\item When $\psi$ is atomic, then $F(\psi) =
  \begin{cases}
    \mathit{true} & \text{if } \psi \text{ is satisfiable in } \Gamma \\
    \mathit{false} & \text{otherwise}
  \end{cases}$
\end{compactitem}

\begin{definition}\label{def:locally-refutable}
  We call a structure~$\Gamma$ \emph{locally refutable} if every
  existential positive sentence~$\Phi$ is true in~$\Gamma$ \emph{if
    and only if} the $\Gamma$-localizer $F(\Phi)$ is logically
  equivalent to \emph{true}.
\end{definition}

\begin{proposition}\label{prop:loc-refut}
  A structure $\Gamma$ is locally refutable if and only if every
  unsatisfiable conjunction of atomic formulas contains an
  unsatisfiable conjunct.
\end{proposition}
\begin{proof}
  First suppose that $\Gamma$ is locally refutable, and let $\phi$ be
  a conjunction of atomic formulas with variables $x_1,\dots,x_n$.
  Then every conjunct of $\phi$ is satisfiable in $\Gamma$ if and only
  if $F(\phi)$ is true. By local refutability of $\Gamma$ this is the
  case if and only if $\exists x_1,\dots,x_n. \phi$ is true in
  $\Gamma$, which shows the claim.

  Now suppose that $\Gamma$ is not locally refutable, that is, there
  is an existential positive sentence $\Phi$ that is false in $\Gamma$
  such that $F(\Phi)$ is true.  Define recursively for each subformula
  $\psi$ of $\Phi$ where $F(\psi)$ is true the formula $T(\psi)$ as
  follows.  If $\psi$ is of the form $\psi_1 \lor \psi_2$, then for
  some $i \in \{1,2\}$ the formula $F(\psi_i)$ must be true, and we
  set $T(\psi)$ to be $T(\psi_i)$.  If $\psi$ is of the form $\psi_1
  \land \psi_2$, then for both $i \in \{1,2\}$ the formula
  $F(\psi_i)$ must be true, and we set $T(\Psi)$ to be $T(\psi_1)
  \land T(\psi_2)$.

  Each conjunct $\phi$ in $T(\Phi)$ is satisfiable in $\Gamma$ since
  $F(\Phi)$ is true.  But since $\Phi$ is false in $\Gamma$, $T(\Phi)$
  must be unsatisfiable.
  \qed
\end{proof}

In Section~\ref{sect:proof}, we will show the following result.

\begin{theorem}\label{thm:main}
  Let~$\Gamma$ be a structure with a finite relational signature
  $\tau$.  If~$\Gamma$ is locally refutable then the problem
  $\ep(\Gamma)$ to decide whether an existential positive sentence is
  true in~$\Gamma$ is in $\logspace$.  If~$\Gamma$ is not locally
  refutable, then $\ep(\Gamma)$ is complete for the class
  $\csp(\Gamma)_\np$ under polynomial-time many-one reductions.
\end{theorem}

In particular, $\ep(\Gamma)$ is in~$\logspace$ or is $\np$-hard (under
deterministic polynomial-time many-one reductions). If~$\Gamma$ is
finite, then $\ep(\Gamma)$ is in~$\logspace$ or $\np$-complete,
because finite domain constraint satisfaction problems are clearly
in~$\np$.  The observation that $\ep(\Gamma)$ is in $\logspace$ or
$\np$-complete has previously been made in~\cite{HerRic-09} and
independently in \cite{Martin-06}.  However, our proof remains the
same for finite domains and is simpler than the previous proofs.

\section{Proof}\label{sect:proof}

Before we prove Theorem~\ref{thm:main}, we start with the following
simpler result.

\begin{theorem}\label{thm:simpler}
  Let~$\Gamma$ be a structure with a finite relational signature
  $\tau$.  If~$\Gamma$ is locally refutable, then the problem
  $\ep(\Gamma)$ to decide whether an existential positive sentence is
  true in~$\Gamma$ is in $\logspace$.  If~$\Gamma$ is not locally
  refutable, then $\ep(\Gamma)$ is $\np$-hard (under polynomial-time
  many-one reductions).
\end{theorem}

To prove Theorem~\ref{thm:simpler}, we need first to prove the
following lemma.

\begin{lemma}\label{lem:tf}
  A structure $\Gamma$ is not locally refutable if and only if there
  are existential positive formulas~$\psi_0$ and~$\psi_1$ with the
  property that
  \begin{compactitem}[-]
  \item $\psi_0$ and $\psi_1$ define non-empty relations
    over~$\Gamma$;
  \item $\psi_0 \land \psi_1$ defines the empty relation
    over~$\Gamma$.
  \end{compactitem}
\end{lemma}
\begin{proof}
  The ``if''-part of the statement is immediate. To show the ``only
  if''-part, suppose that $\Gamma$ is not locally refutable. Then by
  Proposition~\ref{prop:loc-refut} there is an unsatisfiable
  conjunction $\psi$ of satisfiable atomic formulas. Among all such
  formulas $\psi$, let $\psi$ be one of minimal length.  Let $\psi_0$
  be one of the atomic formulas in $\psi$, and let $\psi_1$ be the
  conjunction over the remaining conjuncts in $\psi$.  Since $\psi$
  was chosen to be minimal, the formula $\psi_1$ must be satisfiable.
  By construction $\psi_0$ is also satisfiable and $\psi$ is
  unsatisfiable, which is what we had to show. \qed
\end{proof}

\begin{proofof}{Theorem~\ref{thm:simpler}}
  If $\Gamma$ is locally refutable, then $\ep(\Gamma)$ can be reduced
  to the positive Boolean formula evaluation problem, which is known
  to be $\logspace$-complete. We only have to construct from an
  existential positive sentence $\Phi$ a Boolean formula
  $F:=F_{\Gamma}(\Phi)$ as described before
  Definition~\ref{def:locally-refutable}.  Clearly, this construction
  can be performed with logarithmic work-space. We evaluate $F$, and
  reject if $F$ is false, and accept otherwise.

  If $\Gamma$ is not locally refutable, we show $\np$-hardness of
  $\ep(\Gamma)$ by reduction from 3-SAT.  Let $I$ be a 3-SAT instance.
  We construct an instance $\Phi$ of $\ep(\Gamma)$ as follows.
  Let~$\psi_0$ and~$\psi_1$ be the formulas from Lemma~\ref{lem:tf}
  (suppose they are $d$-ary).  Let $v_1,\dots,v_n$ be the Boolean
  variables in~$I$.  For each $v_i$ we introduce~$d$ new variables
  $\bar x_i = x_i^1,\dots,x_i^d$.  Let~$\Phi$ be the instance of
  $\ep(\Gamma)$ that contains the following conjuncts:
  \begin{itemize}
  \item For each $1 \leq i \leq n$, the formula $\psi_0(\bar x_i) \lor
    \psi_1(\bar x_i)$
  \item For each clause $l_1 \lor l_2 \lor l_3$ in $I$, the formula
    $\psi_{i_1}(\bar x_{j_1}) \lor \psi_{i_2}(\bar x_{j_2}) \lor
    \psi_{i_3}(\bar x_{j_3})$ where $i_p=0$ if $l_p$ equals $\neg
    x_{j_p}$ and $i_p=1$ if $l_p$ equals $x_{j_p}$, for all $p \in
    \{1,2,3\}$.
  \end{itemize}
  It is clear that~$\Phi$ can be computed in deterministic polynomial
  time from~$I$, and that~$\Phi$ is true in~$\Gamma$ if and only if
  $I$ is satisfiable.
  \qed
 \end{proofof}

 Applied to finite relational structures $\Gamma$, we obtain the
 result from~\cite{HerRic-09} and \cite {Martin-06}, that is,
 $\ep(\Gamma)$ is in $\logspace$ if $\Gamma$ is $a$-valid and
 $\np$-complete otherwise.  We prove in the following proposition
 that, over a finite domain $D$, $\Gamma$ is locally refutable if and
 only if it is $a$-valid for an element $a \in D$.

\begin{proposition}\label{prop:locrefdvalid}
  Let $\Gamma$ be a relational structure with a finite domain~$D$.
  Then~$\Gamma$ is locally refutable if and only if it is $a$-valid
  for an element $a \in D$.
\end{proposition}
\begin{proof}
  Suppose that $\Gamma$ is $a$-valid, and let $\Phi$ be an existential
  positive sentence over the signature of $\Gamma$.  To show that
  $\Gamma$ is locally refutable, we only have to show that $\Phi$ is
  true in $\Gamma$ when $F(\Phi)$ is equivalent to true (since the
  other direction holds trivially).  But this follows from the fact
  that if an atomic formula $R(x_1,\dots,x_n)$ is satisfiable in
  $\Gamma$ then in fact this formula can be satisfied by setting all
  variables to $a$.

  For the opposite direction of the statement, let $D =
  \{a_1,\dots,a_n\}$, and suppose that for all $a \in D$ the structure
  $\Gamma$ is not $a$-valid.  That is, for each $a_i \in D$ there
  exists a non-empty relation $R_i$ of arity $r_i$ in $\Gamma$ such
  that $(a_i,\dots,a_i) \notin R$. Let $r$ be $\sum_{i=1}^n r_i$, and
  let $x_1,\dots,x_{rn}$ be distinct variables. Consider the formula
  \begin{eqnarray}
    \psi &=& \bigwedge_{\overline y \in \{x_1,\dots,x_{rn}\}^r}
    R_1(y_1,\dots,y_{r_1}) \land \dots \land R_n(y_{r-r_n+1}, \dots,
    y_{r})
   \label{eq:avalid}
   \end{eqnarray}  
   By the pigeonhole principle, for every mapping $f\colon
   \{x_1,\dots,x_{rn}\} \rightarrow D$ at least $r$ variables are
   mapped to the same value, say to $a_i$. For a vector $\overline y$
   that contains exactly these $r$ variables, for some $l$ there is a
   conjunct $R_i(y_{l+1},\dots,y_{l+r_i})$ in $\psi$; but by
   assumption, $R_i$ does not contain the tuple $(a_i,\dots,a_i)$.
   This shows that $\exists x_1,\dots,x_{rn}. \psi$ is not true
   in~$\Gamma$.  On the other hand, since each relation $R_i$ is
   non-empty, it is clear that the Boolean formula $F(\exists
   x_1,\dots,x_{rn}. \psi)$ is true. Therefore, $\Gamma$ is not
   locally refutable.  \qed
\end{proof}

\begin{remark}\label{rmk:single}
  In the proof of Theorem~\ref{thm:main} it will be convenient to
  assume that $\Gamma$ has a single relation $R$. When we study the
  problem $\csp(\Gamma)$, this is without loss of generality, since we
  can always find a $\csp$ which is deterministic polynomial-time
  equivalent and where the template is of this form: if
  $\Gamma=(D;R_1,\dots,R_n)$ where $R_i$ has arity $r_i$ and is not
  empty, then $\csp(\Gamma)$ is equivalent to $\csp(D;R_1 \times \dots
  \times R_n)$ where $R_1 \times \dots \times R_n$ is the
  $\sum_{i=1}^n r_i$-ary relation defined as the Cartesian product of
  the relations $R_1,\dots,R_n$.  Similarly, $\ep(\Gamma)$ is
  equivalent to $\ep(D;R_1 \times \dots \times R_n)$.
\end{remark}

 \begin{proofof}{Theorem~\ref{thm:main}}
   If~$\Gamma$ is locally refutable then the statement has been shown
   in Theorem~\ref{thm:simpler}.  Suppose that~$\Gamma$ is not locally
   refutable.  To show that $\ep(\Gamma)$ is contained in
   $\csp(\Gamma)_\np$, we construct a non-deterministic Turing
   machine~$T$ which takes as input an instance~$\Phi$ of
   $\ep(\Gamma)$, and which outputs an instance $T(\Phi)$ of
   $\csp(\Gamma)$ as follows.

   On input~$\Phi$ the machine~$T$ proceeds recursively as follows:
   \begin{itemize}
   \item if~$\Phi$ is of the form $\exists x. \phi$ then return
     $\exists x. T(\phi)$;
   \item if~$\Phi$ is of the form $\phi_1 \land \phi_2$ then return
     $T(\phi_1) \land T(\phi_2)$;
   \item if~$\Phi$ is of the form $\phi_1 \lor \phi_2$ then
     non-deterministically return either $T(\phi_1)$ or $T(\phi_2)$;
   \item if~$\Phi$ is of the form $R(x_1,\dots,x_k)$ then return
     $R(x_1,\dots,x_k)$.
   \end{itemize}
   The output of~$T$ can be viewed as an instance of $\csp(\Gamma)$,
   since it can be transformed to a primitive positive sentence (by
   moving all existential quantifiers to the front).  It is clear
   that~$T$ has polynomial running time, and that~$\Phi$ is true
   in~$\Gamma$ if and only if there exists a computation of~$T$
   on~$\Phi$ that computes a sentence that is true in~$\Gamma$.

   We now show that $\ep(\Gamma)$ is hard for $\csp(\Gamma)_\np$ under
   $\leq_m$-reductions.  Let~$L$ be a problem with a non-deterministic
   polynomial-time many-one reduction to $\csp(\Gamma)$, and let~$M$
   be the non-deterministic Turing machine that computes the
   reduction.  We have to construct a deterministic Turing
   machine~$M'$ that computes for any input string~$s$ in polynomial
   time in~$\card s$ an instance~$\Phi$ of $\ep(\Gamma)$ such
   that~$\Phi$ is true in~$\Gamma$ if and only if there exists a
   computation of~$M$ on~$s$ that computes a satisfiable instance of
   $\csp(\Gamma)$.

   Say that the running time of~$M$ on $s$ is in $O(\card s^e)$ for a
   constant $e$.  Hence, there are constants~$s_0$ and~$c$ such that
   for $\card s > s_0$ the running time of $M$ and hence also the
   number of constraints in the input instance of $\csp(\Gamma)$
   produced by the reduction is bounded by $t := c \card s^e$.  The
   non-deterministic computation of~$M$ can be viewed as a
   deterministic computation with access to non-deterministic advice
   bits as shown in \cite{GareyJ-79}.  We also know that for $\card s
   > s_0$, the machine~$M$ can access at most~$t$ non-deterministic
   bits.  If $w$ is a sufficiently long bit-string, we write $M_w$ for
   the deterministic Turing machine obtained from $M$ by using the
   bits in $w$ as the non-deterministic bits, and $M_w(s)$ for the
   instance of $\csp(\Gamma)$ computed by $M_w$ on input $s$.
 
   If $\card s \leq s_0$, then $M'$ returns $\exists \bar x.
   \psi_1(\bar x)$ if there is an $w\in \{0,1\}^*$ such that $M_w(s)$
   is a satisfiable instance of $\csp(\Gamma)$, and $M'$ returns
   $\exists \bar x (\psi_0(\bar x) \land \psi_1(\bar x))$ otherwise
   (i.e., it returns a false instance of $\ep(\Gamma)$; $\psi_0$ and
   $\psi_1$ are defined in Lemma~\ref{lem:tf}).  Since $s_0$ is a
   fixed finite value, $M'$ can perform these computations in constant
   time.

   By Remark~\ref{rmk:single} made above, we can assume without loss
   of generality that~$\Gamma$ has just a single relation~$R$.  Let
   $l$ be the arity of~$R$. Then instances of $\csp(\Gamma)$ with
   variables $x_1,\dots,x_n$ can be encoded as sequences of numbers
   that are represented by binary strings of length $\lceil \log t
   \rceil$ as follows: the $i$-th number $m$ in this sequence
   indicates that the $(((i-1) \bmod l)+1)$-st variable in the
   $(((i-1) \mathbin{\it div} l)+1)$-st constraint is~$x_m$.

   For $\card s>s_0$, we use a construction from the proof of Cook's
   theorem given in~\cite{GareyJ-79}.  In this proof, a computation of
   a non-deterministic Turing machine~$T$ accepting a language $L$ is
   encoded by Boolean variables that represent the state and the
   position of the read-write head of~$T$ at time~$r$, and the content
   of the tape at position~$j$ at time~$r$.  The tape content at
   time~$0$ consists of the input~$x$, written at positions~$1$
   through~$n$, and the non-deterministic advice bit string~$w$,
   written at positions $-1$ through $-\card w$.  The proof
   in~\cite{GareyJ-79} specifies a deterministic polynomial-time
   computable transformation~$f_L$ that computes for a given
   string~$s$ a SAT instance $f_L(s)$ such that there is an accepting
   computation of~$T$ on~$s$ if and only if there is a satisfying
   truth assignment for $f_L(s)$.
  
   In our case, the machine $M$ computes a reduction and thus computes
   an output string.  Recall our binary representation of instances of
   the CSP~$M$ writes on the output tape a sequence of numbers
   represented by binary strings of length $\lceil \log t \rceil$.  It
   is straightforward to modify the transformation $f_L$ given in the
   proof of Theorem 2.1 in~\cite{GareyJ-79} to obtain for all positive
   integers $a,b,c$ where $a \leq t$, $b \leq l$, $c \leq \lceil \log
   t \rceil$, and $d \in \{0,1\}$, a deterministic polynomial-time
   transformation $g^d_{a,b,c}$ that computes for a given string $s$ a
   SAT instance $g^d_{a,b,c}(s)$ with distinguished variables
   $z_1,\dots,z_p$, $p \leq t$ for the non-deterministic bits in the
   computation of $M$ such that the following are equivalent:

   \begin{itemize}
   \item $g^d_{a,b,c}(s)$ has a satisfying assignment where $z_i$ is
     set to $w_i \in \{0,1\}$ for $1 \leq i \leq p$;
   \item the $c$-th bit in the $b$-th variable of the $a$-th
     constraint in $M_w(s)$ equals $d$.
   \end{itemize}

   We use the transformations $g^d_{a,b,c}$ to define~$M'$ as follows.
   The machine $M'$ first computes the formulas $g^d_{a,b,c}(s)$.  For
   every Boolean variable~$v$ in these formulas we introduce a new
   conjunct $\psi_0(\overline x_v) \lor \psi_1(\overline x_v)$ where
   $\overline x_v$ is a $d$-tuple of fresh variables and $\psi_0$ and
   $\psi_1$ are the two formulas defined in Lemma~\ref{lem:tf}.  Then,
   every positive literal $v$ in the original conjuncts of the formula
   is replaced by $\psi_1(\overline x_v)$, and every negative literal
   $l=\neg v$ by $\psi_0(\overline x_v)$.  We then existentially
   quantify over all variables except for $\bar x_{z_1},\dots,\bar
   x_{z_p}$.  Let $\psi^d_{a,b,c}(s)$ denote the resulting existential
   positive formula.  For positive integers $k$ and $i$, we denote as
   $k[i]$ the $i$-th bit in the binary representation of $k$.  Let $n$
   be the total number of variables in the CSP instance $M_w(s)$ (in
   particular, $n \leq t$).  It is clear that the formula
   \begin{displaymath}
     \exists y_1,\dots,y_n,\bar x_{z_1},\dots,\bar x_{z_p}.
     \bigwedge_{1 \leq a, k_1,\dots, k_l \leq t} \left( \left(
         \bigwedge_{b \leq l,c} \psi^{k_b[c]}_{a,b,c}(s) \right)
       \to R(y_{k_1},\dots,y_{k_l}) \right)
   \end{displaymath}
   can be re-written in existential positive form $\Phi$ without
   blow-up: we can replace implications $\alpha \rightarrow \beta$ by
   $\neg \alpha \lor \beta$, and then move the negation to the atomic
   level, where we can remove negation by exchanging the role of
   $\phi_0$ and $\phi_1$. Hence, $\Phi$ can be computed by $M'$ in
   polynomial time.
  
   We claim that the formula $\Phi$ is true in $\Gamma$ if and only if
   there exists a computation of $M$ on $s$ that computes a
   satisfiable instance of $\csp(\Gamma)$.
   To see this, let $w$ be a sufficiently long bit-string such that
   $M_w(s)$ is a satisfiable instance of $\csp(\Gamma)$.  Suppose for
   the sake of notation that the $n$ variables in $M_w(s)$ are the
   variables $y_1,\dots,y_n$.  Let $a_1,\dots,a_n$ be a satisfying
   assignment to those $n$ variables.  Then, if for $1 \leq i \leq n$
   the variable $y_i$ in the formula $\Phi$ is set to $a_i$, and for
   $1 \leq i \leq p$ the variables $\bar x_{z_i}$ are set to a tuple
   that satisfies $\psi_d$ where $d$ is the $i$-th bit in $w$, we
   claim that the inner part of $\Phi$ is true in $\Gamma$.  The
   reason is that, due to the way how we set the variables of the form
   $\bar x_{z_i}$, the precondition $\left( \bigwedge_{b \leq l,c}
     \psi^{k_b[c]}_{a,b,c}(s) \right)$ is true if and only if
   $R(y_{k_1},\dots,y_{k_l})$ is a constraint in $M_w(s)$. Therefore,
   all the atomic formulas of the form $R(y_{k_1},\dots,x_{k_l})$ are
   satisfied due to the way how we set the variables $y_i$, and hence
   $\Phi$ is true in $\Gamma$. It is straightforward to verify that
   the opposite implication holds as well, and this shows the claimed
   equivalence.  \qed
\end{proofof}

\section{Structures With Function Symbols}
\label{sect:functions}

In this section, we briefly discuss the complexity of $\ep(\Gamma)$
when $\Gamma$ might also contain functions.  That is, we assume that
the signature of $\Gamma$ consists of a finite set of relation and
function symbols, and that the input formulas for the problem
$\ep(\Gamma)$ are existential positive first-order formulas over this
signature.  It is easy to see from the proofs in the previous section
that when $\Gamma$ is not locally refutable, then $\ep(\Gamma)$ is
still $\np$-hard (with the same definition of local refutability as
before).

The case when $\Gamma$ is locally refutable becomes more intricate
when $\Gamma$ has functions. We present an example of a locally
refutable structure $\Gamma$ where $\ep(\Gamma)$ is $\np$-hard.  Let
the signature of $\Gamma$ be the structure $(2^{\mathbb N};
\neq,\cap,\cup,c,{\bf 0}, {\bf 1})$ where $\neq$ is the binary
disequality relation, $\cap$ and $\cup$ are binary functions for
intersection and union, respectively, $c$ is a unary function for
complementation, and ${\bf 0},{\bf 1}$ are constants (i.e., $0$-ary
functions) for the empty set and the full set $\mathbb N$,
respectively.

\begin{proposition}
  The structure $(2^{\mathbb N}; \neq,\cap,\cup,c,{\bf 0}, {\bf 1})$
  is locally refutable.
\end{proposition}
\begin{proof}
  By Lemma~\ref{lem:tf} is suffices to show that if~$\Psi$ is a
  conjunction of atomic formulas that are satisfiable in~$\Gamma$,
  then $\Psi$ is satisfiable over $\Gamma$. Since the only relation
  symbol in the structure is $\neq$, every conjunct in $\Psi$ is of
  the form $t_1 \neq t_2$, where $t_1$ and $t_2$ are terms formed by
  variables and the function symbols $\cap$, $\cup$, $c$, $\bf 1$ and
  $\bf 0$.  By Boole's fundamental theorem of Boolean algebras, $t =
  t'$ can be re-written as $t'' = \bf 0$. Therefore, $\Psi$ can be
  written as $t_1 \neq {\bf 0} \land \dots \land t_n \neq {\bf 0}$.
  Since $\Gamma$ is an infinite Boolean algebra, Theorem 5.1
  in~\cite{MO96} shows that if $t_i \neq {\bf 0}$ is satisfiable in
  $\Gamma$ for all $i \leq n$, then $\Psi$ is satisfiable in $\Gamma$
  as well. \qed
\end{proof}

\begin{proposition}
  The problem $\ep(2^{\mathbb N}; \neq,\cap,\cup,c,{\bf 0}, {\bf 1})$
  is $\np$-hard.
\end{proposition}
\begin{proof}
  The proof is by reduction from SAT. Given a Boolean formula $\Psi$
  in CNF with variables $x_1,\dots,x_n$, we replace each conjunction
  in $\Psi$ by $\cap$, each disjunction by $\cup$, and each negation
  by $c$. Let $t$ be the resulting term over the signature
  $\{\cap,\cup,c\}$ and variables $x_1,\dots,x_n$.  It is easy to
  verify that $\exists x_1,\dots,x_n. t \neq \bf 0$ is true in
  $\Gamma$ if and only if $\Psi$ is a satisfiable Boolean formula.
  \qed
\end{proof}

\section{Conclusion}

In this paper, we proved that for an arbitrary (finite or infinite)
relational structure the problem $\ep(\Gamma)$ is in $\logspace$
if~$\Gamma$ is locally refutable, or otherwise complete for the class
$\csp(\Gamma)_\np$ under deterministic polynomial-time many-one
reductions. In particular, if~$\Gamma$ is not locally refutable then
the problem $\ep(\Gamma)$ is $\np$-hard.  Structures with a finite
domain are locally refutable if and only if they are $a$-valid for
some value $a$ of the domain $D$.  Finally, we present an example of a
structure that shows that our result cannot be straightforwardly
extended to structures $\Gamma$ with function symbols, since local
refutability of $\Gamma$ no longer implies that $\ep(\Gamma)$ is in
$\logspace$ when $\Gamma$ contains function symbols.

\section*{Acknowledgment}

We would like to thank V{\'\i}ctor Dalmau for helpful suggestions, and
Moritz M{\"u}ller for the encouragement to study the case where the
structure $\Gamma$ contains function symbols.


\begin{thebibliography}{10}

\bibitem{Vollmer-08}
  N.~Creignou, Ph. G. Kolaitis, and H.~Vollmer, editors.
  \newblock \emph{Complexity   of  Constraints  --- An Overview of
    Current Research Themes}, 
  volume 5250 of Lecture Notes in Computer Science, \emph{Springer Verlag},
  2008.

\bibitem{BodirskyG-08}
  M.~Bodirsky and M.~Grohe.
  \newblock Non-Dichotomies in Constraint Satisfaction Complexity.
  \newblock   \emph{ Proceedings   35th  International   Colloquium  on
    Automata,  Languages   and  Programming  (ICALP   2008),  Part~II,
    Reykjavik (Iceland)}, volume 5126 of Lecture Notes in Computer
  Science, 184--196, 2008.

\bibitem{BHR-09}
  M.~Bodirsky, M.~Hermann and F.~Richoux.
  \newblock  Complexity of Existential Positive First-Order Logic.
  \newblock \emph{Proceedings 5th Conference on Computability in Europe 
    (CiE 2009), Heidelberg (Germany)},
  31--36, 2009.

\bibitem{GareyJ-79}
  M.~R.~Garey and D.~S.~Johnson.
  \newblock  Computers and Intractability:  A Guide  to the  Theory of
  NP-Completeness. 
  \newblock \emph{W.H. Freeman and Co},
  1979.

\bibitem{HerRic-09}
  M.~Hermann and F.~Richoux.
  \newblock  On the Computational Complexity  of Monotone Constraint
  Satisfaction Problems.
  \newblock  \emph{Proceedings  3rd Annual  Workshop on  Algorithms and
    Computation (WALCOM 2009), Kolkata (India)},
  286--297, 2009.

\bibitem{LadnerLS-75}
  R.~E.~Ladner, N.~A.~Lynch and A.~L.~Selman.
  \newblock A Comparison of Polynomial-Time Reducibilities.
  \newblock\emph{Theoretical Computer Science},
  1(2), 103--124, 1975.

\bibitem{MO96}
  K.~Marriott and M.~Odersky.
  \newblock Negative boolean constraints.
  \newblock \emph{Theoretical Computer Science}, 160(1{\&}2), 365--380, 1996.

\bibitem{Martin-06}
  B.~Martin.
  \newblock Dichotomies and Duality in First-order Model Checking Problems.
  \newblock \emph{CoRR abs/cs/0609022},
  2006.

\bibitem{Martin-08}
  B.~Martin.
  \newblock First-Order Model Checking Problems Parameterized by the Model.
  \newblock \emph{Proceedings 4th Conference on Computability in Europe
    (CiE 2008), Athens (Greece)}, 
  417--427, 2008.

\end{thebibliography}
\end{document}